\documentclass[12pt]{article}

\usepackage{graphicx}

\usepackage{amsmath}
\usepackage{amssymb}
\allowdisplaybreaks

\begin{document}

\baselineskip=17.5pt plus 0.2pt minus 0.1pt

%%%%%% private definition %%%%%%%
\renewcommand{\theequation}{\arabic{equation}}
\renewcommand{\thefootnote}{\fnsymbol{footnote}}
\makeatletter
%\@addtoreset{equation}{section}
\def\CR{\nonumber \\}
\def\be{\begin{equation}}
\def\ee{\end{equation}}
\def\bea{\begin{eqnarray}}
\def\eea{\end{eqnarray}}
\def\beal{\be\begin{aligned}}
\def\eeal{\end{aligned}\ee}
\def\eq#1{(\ref{#1})}
\def\la{\langle}
\def\ra{\rangle}
\def\hyp{\hbox{-}}
\def\ul{\underline}
%%%%%%%%%%%%%%%%%%%%%%%%%%%%%%%%%

\begin{titlepage}
\title{\hfill\parbox{4cm}{ \normalsize YITP-12-12}\\
\vspace{1cm} Uniqueness of canonical tensor model with local time}
\author{Naoki {\sc Sasakura}
\thanks{\tt sasakura@yukawa.kyoto-u.ac.jp}
\\[15pt]
{\it Yukawa Institute for Theoretical Physics, Kyoto University,}\\
{\it Kyoto 606-8502, Japan}}
\date{}%\normalsize}
\maketitle
\thispagestyle{empty}
\begin{abstract}
\normalsize
Canonical formalism of the rank-three tensor model has recently been proposed, in which
``local" time is consistently incorporated by a set of first class constraints.
By brute-force analysis, this paper shows that there exist only two forms of a
Hamiltonian constraint which satisfies the following assumptions: 
(i) A Hamiltonian constraint has one index. 
(ii) The kinematical symmetry is given by an orthogonal group.
(iii) A consistent first class constraint algebra is formed by a Hamiltonian constraint and the generators of the kinematical symmetry. 
(iv) A Hamiltonian constraint is invariant under time reversal transformation. 
(v) A Hamiltonian constraint is an at most cubic polynomial function of canonical variables. 
(vi) There are no disconnected terms in a constraint algebra.
The two forms are the same except for a slight difference in index contractions.
The Hamiltonian constraint which was obtained in the previous paper and behaved oddly under time 
reversal symmetry can actually be transformed to one of them
by a canonical change of variables.  
The two-fold uniqueness is shown up to the potential ambiguity of adding terms which 
vanish in the limit of pure gravitational physics.
\end{abstract}
\end{titlepage}

\section{Introduction}
\label{sec:introduction}
Though quantum gravity has not yet fully been constructed, 
theoretical arguments based on the combination of the general relativity and quantum mechanics suggest that 
quantum gravitational fluctuations destroy precise spacetime measurements
around the Planck energy \cite{Garay:1994en}. 
This prompts the quest for a new quantum notion of spacetime in place 
of the classical one, which is based on 
continuous and smooth manifolds.
From this perspective,
the classical spacetime manifold is merely an infrared effective notion which emerges from underlying 
fundamental dynamics \cite{Sindoni:2011ej}.

An interesting candidate of such a  quantum notion is the fuzzy space, which
describes a space with an algebra of functions on it\footnote{
Noncommutative spaces \cite{Connes:1994yd,Madore:1991bw} are the special classes of fuzzy spaces, 
which are described by noncommutative associative algebra of functions.
Nonassociative spaces \cite{deMedeiros:2004wb}-\cite{Sasai:2006ua} are also of physical interest.} \cite{Connes:1994yd}-\cite{Sasai:2006ua}.
As discussed in \cite{Sasakura:2005js}-\cite{Sasakura:2011qg},  
the dynamics of fuzzy spaces can be formulated
as the rank-three tensor models,
which have a rank-three tensor as their only dynamical variable.
Then the rank-three tensor models can be regarded as a kind of quantum gravity.

In fact, tensor models have originally been proposed as a formulation of simplicial quantum 
gravity in dimensions higher than two \cite{Ambjorn:1990ge,Sasakura:1990fs,Godfrey:1990dt}.
The tensor models have later been generalized to describe topological lattice theories
\cite{Boulatov:1992vp,Ooguri:1992eb} and 
the loop quantum gravity \cite{DePietri:1999bx,Freidel:2005qe,Oriti:2011jm} 
by considering Lie-group valued indices.
Interesting recent developments are the advent of the colored tensor models \cite{Gurau:2009tw} 
and the subsequent discussions \cite{Bonzom:2012hw}-\cite{Gurau:2009tz}, which have presented 
improved formulations of tensor models.
There have also been semi-classical studies of the rank-three tensor models by the present author
\cite{Sasakura:2006pq}-\cite{Sasakura:2010rb},
based on the interpretation in the previous paragraph. These semi-classical works have numerically 
shown the emergence of the (Euclidean) general relativity in the perturbations 
around the backgrounds representing various dimensional fuzzy flat spaces, 
which are classical solutions to 
certain fine-tuned rank-three tensor models.

The developments so far in tensor models have basically been dealing with the Euclidean cases.
While field theories in flat Minkowski spacetimes can be connected to Euclidean ones 
by analytical continuation as a standard procedure, 
it seems a subtle problem whether this is also true in quantum gravity. 
In fact, the results of Causal Dynamical Triangulation suggest otherwise \cite{Ambjorn:2010ce}. 
Another serious problem in the Euclidean tensor models
is that their actions can freely be chosen as long as they respect the kinematical symmetries,
and the tensor models have no predictive powers for possible future observations, 
while there may be chances for universality to save the situation 
\cite{BenGeloun:2012pu,BenGeloun:2011rc,Geloun:2011cy}.

In the previous paper \cite{Sasakura:2011sq},
to overcome those problems in the Euclidean tensor models,
the present author discussed the canonical formalism of the rank-three
tensor model with ``local" time. As in the canonical formalism of the general relativity 
\cite{Arnowitt:1962hi,DeWitt:1967yk,Hojman:1976vp},
the form of the Hamiltonian constraint is strongly constrained by the requirement of the 
algebraic closure of the first class constraints. 
The previous paper has certainly provided a consistent Hamiltonian constraint, 
but it was not clarified 
whether it was unique or not. Moreover, it seemed problematic that
the Hamiltonian constraint had a form which behaved oddly under the time reversal transformation. 

The purpose of the present paper is to write down all the allowed forms of a Hamiltonian constraint
under the following physically reasonable assumptions:
(i) A Hamiltonian constraint has one index. (ii) The kinematical symmetry is 
given by an orthogonal group.
(iii) A consistent first class constraint algebra is formed by a Hamiltonian constraint and 
the generators of the kinematical symmetry. 
(iv) A Hamiltonian constraint is invariant under time reversal transformation. 
(v) A Hamiltonian constraint is an at most cubic polynomial function of canonical variables. 
(vi) There are no disconnected terms in a constraint algebra.

The discussions will be based on brute-force analysis.
The general form of a Hamiltonian constraint respecting the assumptions except for the algebraic ones
in the above 
will be written down, and then the algebraic assumptions will be imposed by explicitly computing 
the Poisson brackets for the general form.
It will be shown that only two forms of a Hamiltonian
constraint, which have a slight difference in  
index contractions, satisfy the assumptions.
It will also be found that, after a canonical change of variables, 
the Hamiltonian constraint which was obtained in the previous paper
is indeed identical to one of them.
The two-fold uniqueness above will be shown up to the potential ambiguity of adding terms which 
will vanish in the limit of pure gravitational physics.

This paper is organized as follows. In Section \ref{sec:graph}, 
graphical representation  will be introduced to 
efficiently carry out the computations, which would otherwise become very much cumbersome. 
It will be explained that the graphical representation introduced in this section is not completely 
precise, but will give some necessary conditions, which will turn out to 
drastically reduce the possibilities of a Hamiltonian constraint.
In Section \ref{sec:generalham},  some general properties of a Hamiltonian constraint
will be discussed. Then
all the possible terms of a Hamilton constraint 
respecting the above assumptions except for 
the algebraic ones will be written down by using the graphical representation.
In Section \ref{sec:poisson},
the Poisson bracket of the Hamiltonian constraints with the general terms will be computed
by using the graphs.
In Section \ref{sec:solution}, the algebraic assumptions will be imposed, and the 
solutions satisfying the necessary conditions will be obtained.
In Section \ref{sec:order}, to find the full solutions, signatures will be introduced to 
the graphical representation to precisely represent the Poisson bracket, 
and the complete conditions will be obtained.
It will be shown that there exist only two forms for a Hamiltonian constraint.
Section \ref{sec:summary} will be devoted to the summary and future prospects.

\section{Graphical representation}
\label{sec:graph}
The canonical variables of the rank-three tensor models are assumed to be given by 
$M_{abc} \ (a,b,c=1,2,\ldots,N)$ and its conjugate momentum $\pi_{abc}$. They are assumed to
satisfy the generalized Hermiticity condition,
\beal
\label{eq:hermiticity}
&M_{abc}=M_{bca}=M_{cab}=M^*_{bac}=M^*_{acb}=M^*_{cba}, \\
&\pi_{abc}=\pi_{bca}=\pi_{cab}=\pi^*_{bac}=\pi^*_{acb}=\pi^*_{cba},
\eeal
where $*$ denotes the complex conjugation. The Poisson brackets of the canonical variables are assumed to be given by
\beal
&\{M_{abc},\pi_{def}\}=\delta_{abc,def}\equiv\delta_{ad}\delta_{be}\delta_{cf}+
\delta_{ae}\delta_{bf}\delta_{cd}+\delta_{af}\delta_{bd}\delta_{ce},\\
&\{M_{abc},M_{def}\}=\{\pi_{abc},\pi_{def}\}=0,
\label{eq:canonicalpoisson}
\eeal
which respect the generalized Hermiticity \eq{eq:hermiticity}.

\begin{figure}
\centering
\includegraphics[width=3cm]{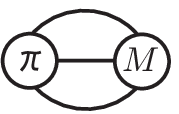}
\caption{The graphical representation of $c_1 \pi_{abc}M_{abc}+c_2 \pi_{abc}M_{bac}$ with $c_1+c_2=1$.}
\label{fig:pim}
\end{figure}
The fact that the canonical variables have three indices tends to make computations cumbersome
and inefficient. 
To avoid this,
let me introduce graphical representation as follows. 
As shown in an example in Figure \ref{fig:pim},
a blob represents $M_{abc}$ or $\pi_{abc}$, and connected lines represent contractions of
the indices. While the Hermiticity condition \eq{eq:hermiticity} assures the invariance under
the cyclic rotations of the indices, which correspond to the rotations of a blob, 
the relative orders of the 
indices, such as the distinction between $\pi_{abc}M_{abc}$ and $\pi_{abc}M_{bac}$, are relevant. 
In the graphical representation,
this kind of interchange of the order of the indices would generate crossing of the lines.
In Sections \ref{sec:generalham},
\ref{sec:poisson} and \ref{sec:solution},
to avoid such entanglement of the lines,  
it will be assumed that 
the graphical representation does not care the order: the graph in Figure \ref{fig:pim},
for example, represents either $\pi_{abc}M_{abc}$ or $\pi_{abc}M_{bac}$, or even a linear combination
of them, the coefficients of which are assumed to add up to 1.

In general, this ambiguous treatment of the order of the indices will generate 
ambiguity in the computations of the Poisson brackets. 
As an illustration, let me consider 
the following simple computation of a Poisson bracket,  
\be
\{c_1 \pi_{abc}M_{abc}+c_2 \pi_{abc}M_{bac}, \pi_{abc}\pi_{abc}\}=
6 c_1 \pi_{abc}\pi_{abc}+6 c_2 \pi_{abc}\pi_{bac}.
\label{eq:comAB}
\ee
Let me assume that a graph does not care the order of the indices so that each value of $c_i$ is not
determined, but the sum is fixed by $c_1+c_2=1$.
Though the result in the right-hand side of \eq{eq:comAB} is certainly ambiguous, 
this ambiguity can be deleted
by assuming $\pi_{abc}$ (and also $M_{abc}$ in general) to be real. This is because, from the
reality assumption and the Hermiticity \eq{eq:hermiticity},
$\pi_{abc}$ (and $M_{abc}$) become totally symmetric with respect to the indices, and 
the right-hand side of \eq{eq:comAB} adds up to $6(c_1+c_2)\pi_{abc}\pi_{abc}$,
which is not ambiguous.
This trick of imposing the reality assumption on the results of Poisson brackets 
to obtain unambiguous results
is not always applicable, but can be applied to simple graphs, and will drastically simplify 
the discussions of this paper.

Thus, in Sections \ref{sec:generalham},
\ref{sec:poisson} and \ref{sec:solution}, I will assume the reality of $M_{abc}$ and $\pi_{abc}$
for the results of Poisson brackets. In fact, one can easily check, for each
of the following computations, 
that the results are not ambiguous, if the reality assumption is imposed,
even though a graph does not care the order of the indices.
However, if this reality assumption is taken, only 
some necessary conditions for the algebraic consistency will be obtained,
since the algebraic consistency is considered only on the slice of the real values of 
the canonical variables.
Therefore, after the possibilities of a Hamiltonian constraint have drastically been 
reduced by the necessary conditions,
the complete treatment which cares the order of the indices will follow in Section \ref{sec:order}.

\section{The general form of a Hamiltonian constraint}
\label{sec:generalham}
The kinematical symmetry of the rank-three tensor model is assumed to be given by the orthogonal group $O(N)$,
which is represented by the vector representation on the indices of $M_{abc}$ and $\pi_{abc}$. 
With the canonical variables, the generators are expressed as
\be
{\cal D}_{ab}=\frac{1}{2} \left( \pi_{acd} M_{bcd}-\pi_{bcd} M_{acd}\right).
\label{eq:expd}
\ee
They satisfy
\be
\{D(V^1),D(V^2)\}=D([V^1,V^2]),
\label{eq:poissondd}
\ee
where 
\be
D(V)\equiv V_{ab} {\cal  D}_{ab}
\ee
with an anti-symmetric real matrix $V_{ab}$, and $[\ , \ ]$ denotes a matrix commutator. 

As discussed in the previous paper \cite{Sasakura:2011sq}, a Hamiltonian constraint 
should have an index as ${\cal H}_a$.
It is also assumed that ${\cal H}_a$ is a polynomial function of the canonical variables,
whose indices should be contracted in pairs except for the one corresponding to the index $a$ of 
${\cal H}_a$. Then, since the inner indices are contracted to be 
invariant under the orthogonal group symmetry,
the group transformation applies only to the index $a$ of ${\cal H}_a$, and therefore 
\be
\{D(V),H(T)\}=H(VT)
\label{eq:poissondh}
\ee
is satisfied, where
\be
H(T)\equiv T_a {\cal H}_a
\ee
with a real vector $T_a$.

\begin{figure}
\centering
\includegraphics[width=10cm]{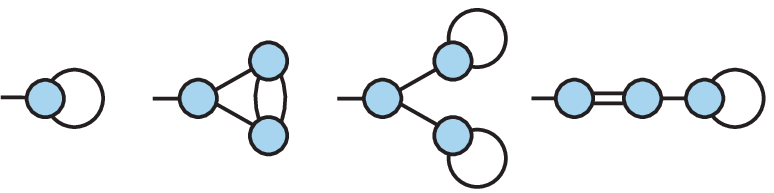}
\caption{The topological structures of the possible terms in ${\cal H}_a$.
The blobs represent $\pi_{abc}$ or $M_{abc}$.}
\label{fig:topology}
\end{figure}
Another assumption is that the terms which compose ${\cal H}_a$ be represented by connected graphs.
This assumption is physically required, because, otherwise, the dynamics of the tensor models 
would become non-local on an emergent space. 
By also assuming ${\cal H}_a$ be at most cubic in canonical variables, the possible topological 
structures of the graphs representing the terms in ${\cal H}_a$ can be summarized as in Figure
\ref{fig:topology}.
\begin{figure}
\centering
\includegraphics[width=10cm]{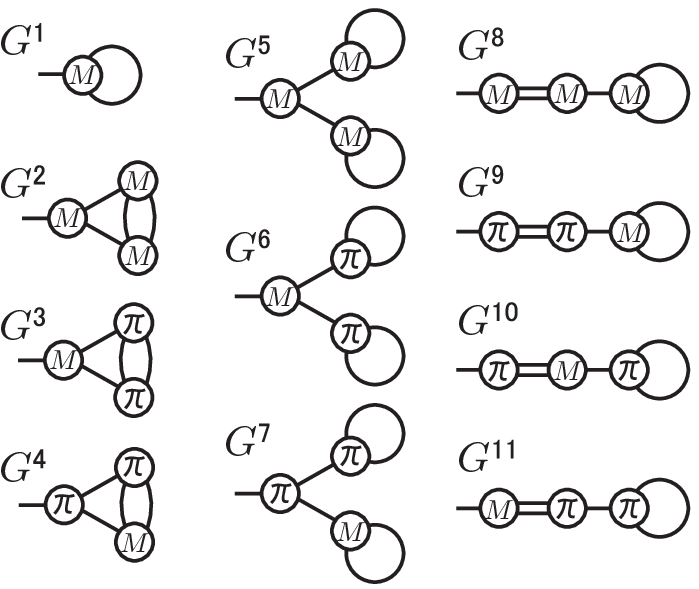}
\caption{All the possible terms in ${\cal H}_a$ which respect the time reversal symmetry.}
\label{fig:possible}
\end{figure}

The assumption of the time reversal symmetry, that ${\cal H}_a$ be invariant under 
$\pi_{abc} \rightarrow -\pi_{abc}$, requires that there are even numbers of $\pi_{abc}$ 
in each term.
Then, by ignoring the index ordering as explained in the previous section, all the possible
terms in ${\cal H}_a$ can be listed as in Figure~\ref{fig:possible}.

\section{Poisson bracket of Hamiltonian constraints}
\label{sec:poisson}
Based on the discussions in the previous section, 
the general form of a Hamiltonian constraint can be expressed as 
\be
{\cal H}_a = \sum_{i=1}^{11}  d_i\, G_{a}^i,
\label{eq:generalform}
\ee
where $d_i$ are real numerical coefficients, and $G^i_a$ are the terms represented by the graphs
in Figure~\ref{fig:possible}.
In this section, the Poisson bracket of the Hamiltonian constraints in the general form
\eq{eq:generalform} will be computed.

Let me introduce the notion of $\pi$-degree, which counts the number of $\pi_{abc}$ in each term. 
For example, the graphs $G^1$ and $G^3$ have the $\pi$-degree 0 and 2, respectively.
This $\pi$-degree
is additive in multiplication, and a Poisson bracket can be considered to have the $\pi$-degree -1,
since the number of $\pi_{abc}$ is reduced by one by computing a non-vanishing Poisson bracket.
Since the $\pi$-degree is a conserved quantity with this assignment, and ${\cal H}_a$ is composed of
the terms with the $\pi$-degree 0 or 2, 
the Poisson bracket of the Hamiltonian constraints is given by
\be
\label{eq:hhdegree}
\{ H(T^1),H(T^2)\}=\hbox{(terms with the $\pi$-degree 3)}+\hbox{(terms with the $\pi$-degree 1)},   
\ee
where the former terms in the right-hand side come from the Poisson brackets 
of the terms with the $\pi$-degree 2,
while the latter terms from those of the terms with the $\pi$-degree 2 and 0.
Therefore, one can separately discuss the closure condition of the constraint algebra
by classifying the terms according to the $\pi$-degree.
So let me first consider the terms with the $\pi$-degree 2 in ${\cal H}_a$
by putting $d_1=d_2=d_5=d_8=0$ to suppress the terms with the $\pi$-degree 0.
The allowed values of $d_1,d_2,d_5,d_8$ will be discussed later in Sections \ref{sec:solution}
and \ref{sec:order}.

\begin{figure}
\centering
\includegraphics[width=10cm]{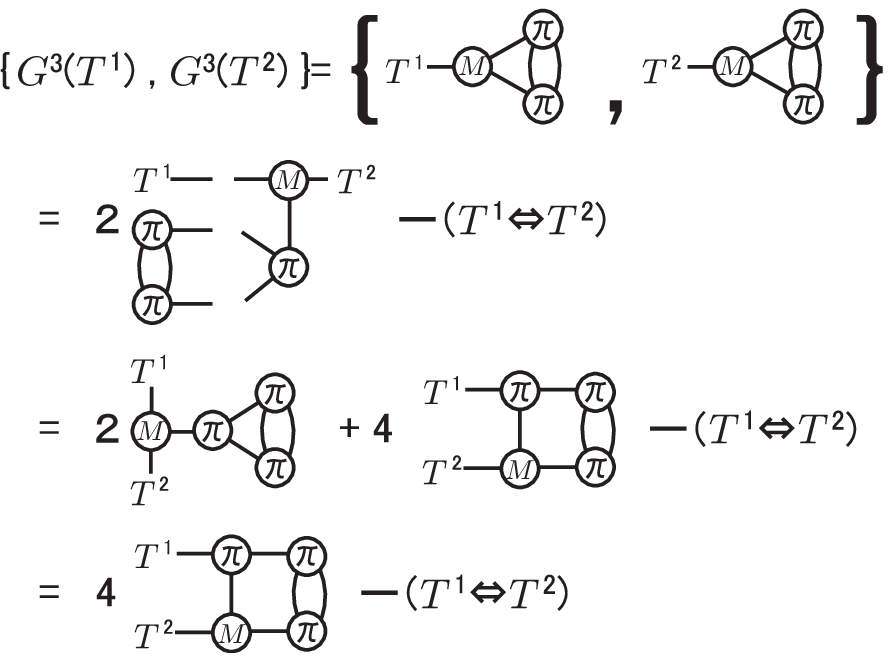}
\caption{The computation of $\{G^3(T^1),G^3(T^2)\}$ by the graphs.}
\label{fig:g3g3}
\end{figure}
The computations of the Poisson brackets can efficiently be carried out by using the graphical
representation.
As an example, the computation of $\{G^3(T^1),G^3(T^2)\}$, where $G^i(T)\equiv T_a G^i_a$,
 is illustrated in Figure \ref{fig:g3g3}.
From the first to the second line of the figure, 
pairs of $M_{abc}$ and $\pi_{abc}$ are deleted from the graphs, either one from
each graph, and, in the third line, the pair of the open graphs in the second line 
are connected by gluing the open lines
in all the possible ways without caring the index ordering. 
From the third to the last line, the graph symmetric under the 
interchange $T^1 \leftrightarrow T^2$ has been canceled by the correspondence
in $-(T^1\leftrightarrow T^2)$ in the figure\footnote{
Note that, as explained in Section \ref{sec:graph}, $M_{abc}$ and $\pi_{abc}$
are totally symmetric with respect to the indices under the reality assumption.}.
In the computation, it is important to take correctly into account the multiplicity and 
also the signature
coming from which of $\{M_{abc},\pi_{def}\}$ or $\{\pi_{abc},M_{def}\}$ is computed.  

\begin{figure}
\centering
\includegraphics[width=13cm]{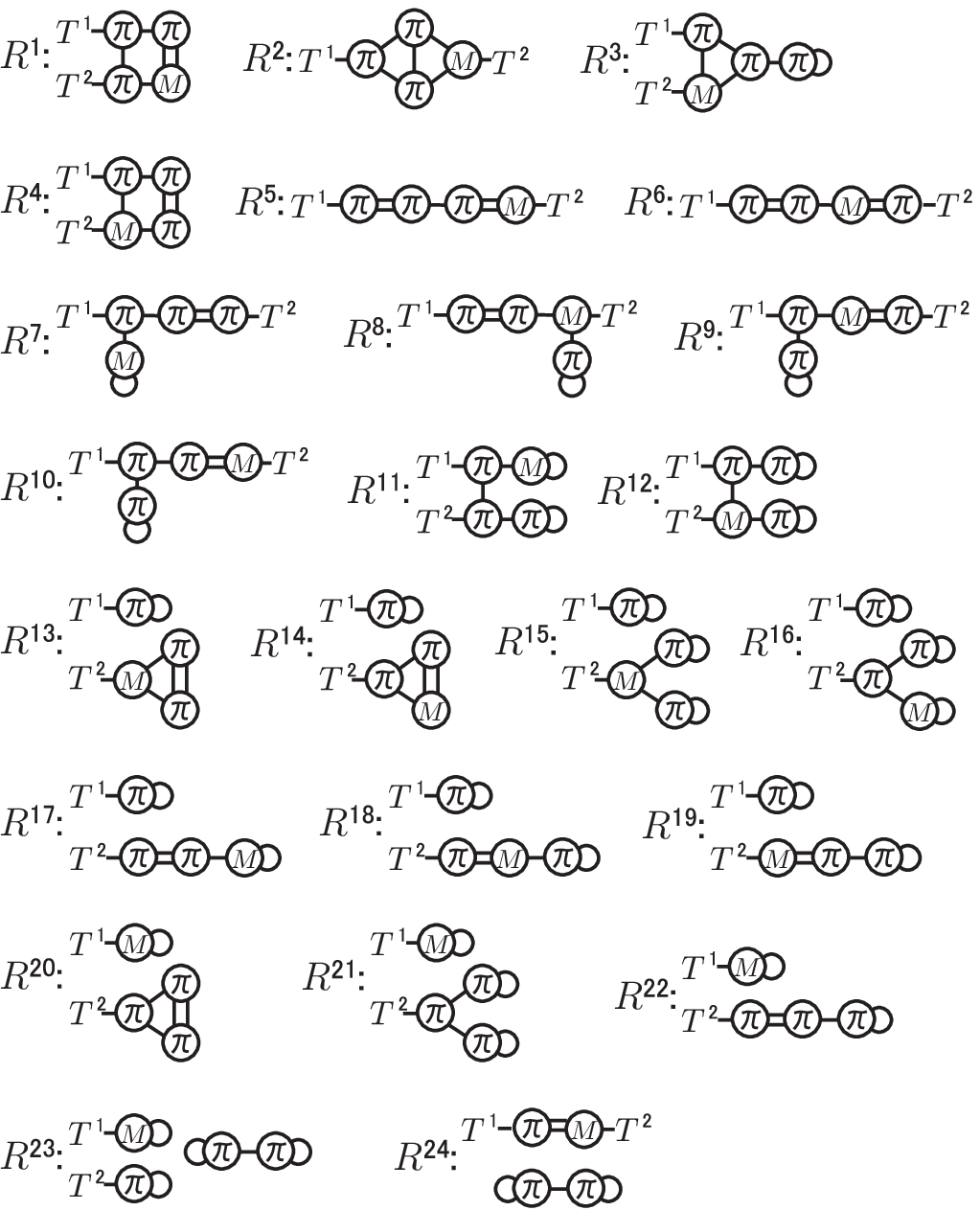}
\caption{All the graphs which are generated from $\{ H(T^1),H(T^2)\}$.}
\label{fig:p3result}
\end{figure} 
The Poisson brackets of the other graphs can be computed in the same manner. 
The result is
\beal
\{&H(T^1),H(T^2)\}=
(d_4)^2  R^1 + 4 d_3 d_4 R^2 + (4 d_3 d_{11} + 4 d_3 d_{10}) R^3 + 4 (d_3 )^2 R^4 \\
 &+(-d_3 d_4 + 2 d_4 d_{11} + d_4 d_9 + 2 d_3 d_{11} + 4 d_3 d_9 + (N + 2) d_9 d_{11}) R^5 \\
 &+(-d_3 d_4 + 2 d_4 d_{10} + 3 d_4 d_9 + 2 d_3 d_{10} + (N + 2) d_9 d_{10})R^6 \\
 &+(2 d_3 d_9 - 4 (d_9)^2 - 2 d_4 d_7 - 2 d_3 d_7 - (N + 2) d_7 d_9) R^7\\  
&+(2 d_3 d_9 + 2 d_9 d_{11} + 2 d_9 d_{10} + 4 d_4 d_6 + 4 d_3 d_6 + 
    2 (N + 2) d_6 d_9) R^8  \\
    &+
 (2 d_{10} d_{11} + 2 (d_{10})^2  + 3 d_4 d_7 + (N + 2) d_7 d_{10})R^9\\
 &  +
(2 (d_{11})^2 + 2 d_{10} d_{11} + d_4 d_7 + 4 d_3 d_7 + (N + 2) d_7 d_{11})  R^{10}\\
& +
 (2 d_9 d_{11} + 2 d_9 d_{10} - (N + 2) (d_7)^2  - 2 d_4 d_7 - 2 d_7 d_{11} - 
    2 d_7 d_{10} - 4 d_7 d_9)  R^{11}\\
    &+
 (2 (N + 2) d_6 d_7 + 2 d_3 d_7 + 2 d_7 d_{11} + 2 d_7 d_{10} + 4 d_3 d_6 + 
    4 d_6 d_{11} + 4 d_6 d_{10}) R^{12}\\
    & +
  d_3 d_{11} R^{13}-d_4 d_{10} R^{14} +
(4 (d_6)^2  + d_6 d_{11}) R^{15}\\
& +
 (-(d_7)^2  + 2 d_6 d_7 + d_7 d_{11} - d_7 d_9 - 2 d_6 d_9) R^{16}+
  (-d_9 d_{10} - (d_9)^2 + d_3 d_7 - d_7 d_9)R^{17}\\
  & +
 (-d_4 d_6 + 2 d_6 d_{10}) R^{18} +
 ((d_{11})^2  + d_{10} d_{11} - d_4 d_6 + 2 d_6 d_{11}) R^{19} +
 d_3 d_9 R^{20} \\
 &  +
  (-(d_7)^2  + d_6 d_9)R^{21} +
 (d_3 d_7 - d_7 d_9)R^{22}  +
 d_6 d_7 R^{23} +
((d_{11})^2 + d_6 d_{11} - d_6 d_{10}) R^{24}\\
&-(T^1 \leftrightarrow T^2), 
\label{eq:hhfinal}
\eeal
where $R^i$'s are defined in Figure \ref{fig:p3result}. 

\section{Solutions to the consistency of the constraint algebra}
\label{sec:solution}
An assumption of this paper is that ${\cal H}_a$ and ${\cal D}_{ab}$
form a consistent first class constraint algebra.  
From \eq{eq:poissondd} and \eq{eq:poissondh}, 
the Poisson brackets containing ${\cal D}_{ab}$ generally satisfy this assumption.
On the other hand, 
the Poisson bracket of $H(T)$'s obtained in \eq{eq:hhfinal} 
has many unwanted terms and the requirement of the algebraic consistency will strongly restrict 
the allowed values of $d_i$.  

First of all, the assumption of the graphical connectivity of the terms existing in the algebra requires that
the terms represented by $R^i\ (i=13,14,\ldots,24)$
should not appear in \eq{eq:hhfinal}, since these graphs are disconnected as listed in Figure \ref{fig:p3result}.

Next, as one can see in Figure \ref{fig:p3result},   
$R^1-(T^1\leftrightarrow T^2)$ and $R^{24}-(T^1\leftrightarrow T^2)$  
contain ${\cal D}_{ab}$ as their parts. 
This is also true for $R^5-R^6$ and $R^9-R^{10}$.
Therefore these terms in \eq{eq:hhfinal}
do not violate the consistency of the constraint algebra, and are allowed.
 
The other graphs or combinations of the graphs in Figure \ref{fig:p3result}
do not contain the same structures as the graphs of ${\cal H}_a$ in Figure \ref{fig:possible} or of ${\cal D}_{ab}$.
Therefore all the coefficients of these terms in \eq{eq:hhfinal} must vanish.
The conditions for this vanishing to hold for general $N$ are given by
\beal
d_3 d_4&=0 ,\\ 
d_3 (d_{10} + d_{11})&=0 ,\\
(d_3 )^2&=0 ,\\
-d_3 d_4 + d_4 (d_{10}+d_{11}) +2 d_4 d_9 +  d_3 (d_{10}+d_{11}) + 2 d_3 d_9&=0,\\
d_9 (d_{10}+d_{11})&=0 ,\\
d_3 d_9 - 2 (d_9)^2 -  d_4 d_7 -  d_3 d_7&=0,\\
d_7 d_9&=0,\\  
d_3 d_9 +  d_9( d_{10} + d_{11}) + 2 d_4 d_6 + 2 d_3 d_6&=0,\\
d_6 d_9&=0  ,\\
(d_{10}+ d_{11})^2 + 2 (d_3+ d_4 ) d_7&=0,\\
d_7 (d_{10}+d_{11})&=0,\\
( d_9-d_7) (d_{10} + d_{11})  -  d_4 d_7 - 2 d_7 d_9&=0,\\ 
(d_7)^2&=0, \\
 d_3 d_7 +  d_7 (d_{10} + d_{11}) + 2 d_3 d_6 + 
    2 d_6 (d_{10} + d_{11})&=0 ,\\
d_6 d_7&=0.
\label{eq:othereq}
\eeal 
Here I have taken into account the allowance explained in the previous paragraph.
These equations determine
\begin{align}
d_3=d_7=d_9=d_{10}+d_{11}=0,
\label{eq:dszero}
\end{align}
while there exist two non-vanishing cases for $d_4$ and $d_6$ as\footnote{
$d_4=d_6=0$ is not appropriate, since then the Hamiltonian constraint contains only the terms proportional to ${\cal D}_{ab}$.}
\begin{align}
\hbox{(i)}&\ d_4\neq 0,\ d_6=0, 
\label{eq:casei}
\\
\hbox{(ii)}&\ d_4=0,\ d_6\neq 0.
\label{eq:caseii}
\end{align}

The case (ii) in \eq{eq:caseii} is not appropriate, since one can easily show that this case 
contradicts
the absence of the disconnected terms explained above. 

On the other hand, for the case (i), by substituting \eq{eq:casei} into \eq{eq:hhfinal}, 
one finds that, for the absence of the disconnected terms, $d_{10}=d_{11}=0$ 
is also required. Then the Poisson bracket \eq{eq:hhfinal} becomes $R^1-(T^1 \leftrightarrow T^2)$.
Since $R^1-(T^1 \leftrightarrow T^2)$ contains ${\cal D}_{ab}$ as its part, the case (i) is the primary candidate 
for a consistent constraint algebra.
So, the only physically consistent solution is the case (i), 
and the terms with the $\pi$-degree 2 in ${\cal H}_a$ are exhausted by $G^4$.  

Let me next study the terms with the $\pi$-degree 1 
in the right-hand side of \eq{eq:hhdegree}.
These terms come from the Poisson brackets of $G^4$ 
and the graphs, $G^1,G^2,G^5,G^8$, with the $\pi$-degree 0. 
Since $G^1$ and the graphs, $G^2,G^5,G^8$, 
have different degrees of the canonical variables, 
the two kinds of graphs can be discussed separately.
Let me consider only $G^2,G^5,G^8$ in the following discussions,
leaving $G^1$ for later discussions in Section \ref{sec:order}. 

\begin{figure}
\centering
\includegraphics[width=15cm]{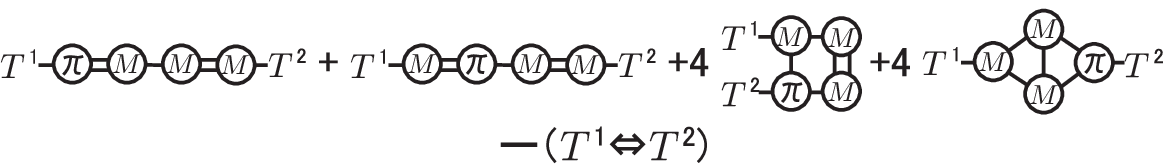}
\caption{The result of $\{G^4(T^1),G^2(T^2)\}+\{G^2(T^1),G^4(T^2)\}$.}
\label{fig:g4g2}
\end{figure}
\begin{figure}
\centering
\includegraphics[width=10cm]{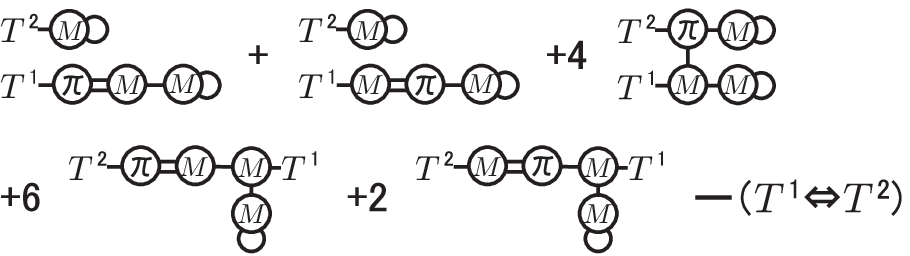}
\caption{The result of $\{G^4(T^1),G^5(T^2)\}+\{G^5(T^1),G^4(T^2)\}$.}
\label{fig:g4g5}
\end{figure}\begin{figure}
\centering
\includegraphics[width=13cm]{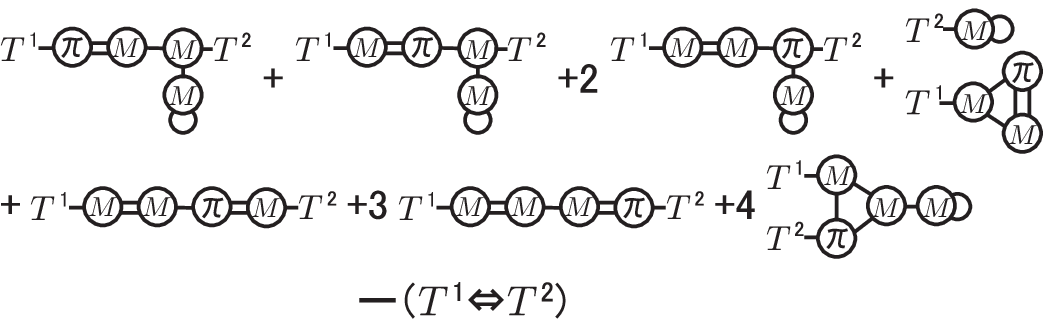}
\caption{The result of $\{G^4(T^1),G^8(T^2)\}+\{G^8(T^1),G^4(T^2)\}$.}
\label{fig:g4g8}
\end{figure}
The Poisson brackets, 
$\{G^4(T^1),G^i(T^2)\}+\{G^i(T^1),G^4(T^2)\}\ (i=2,5,8)$,
can be computed graphically as before, 
and the results are given in Figures \ref{fig:g4g2}, \ref{fig:g4g5} and \ref{fig:g4g8}.
Since each of these figures contains at least one graph, which is not contained in the 
other figures and contains no parts identical to ${\cal D}_{ab}$, the only solution to the 
algebraic consistency is the vanishing one, $d_2=d_5=d_8=0$.

\section{Computations respecting the order of the indices}
\label{sec:order}
\begin{figure}
\centering
\includegraphics[width=10cm]{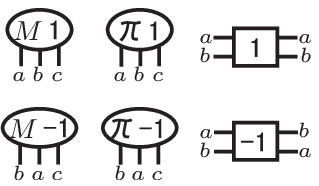}
\caption{In the leftmost two figures, $M_{abc}$ is represented in two ways, in which the order
of the indices depends on the signature. 
In the central two figures, $\pi_{abc}$ is also represented in the two ways. 
In the rightmost upper figure, 
the left couple of the lines are connected to the right ones in parallel, while, in the rightmost 
lower figure, they are connected after being twisted.} 
\label{fig:orientgraph}
\end{figure}

\begin{figure}
\centering
\includegraphics[width=10cm]{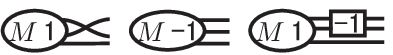}
\caption{The new graphs can be used to unwind the entangled lines. The three graphs in the 
figure have the same connectivity of the lines.} 
\label{fig:unwind}
\end{figure}

The previous section concludes that the Hamiltonian constraint can only have the terms represented
by $G^4$ and possibly by $G^1$.  
As explained in Section \ref{sec:graph}, the graphical representation in Sections \ref{sec:generalham},
\ref{sec:poisson} and \ref{sec:solution}  
is not supposed to unambiguously specify the index contractions of the canonical variables, 
and cannot fully determine a Hamiltonian constraint.
To obtain the full solutions, 
one has to keep precise track of the connections of the lines in a graph, which will generally generate 
a complicated entangled graph. To avoid this entanglement,
let me introduce signatures $\pm 1$ to the graphs as in Figure \ref{fig:orientgraph}.
These new graphs unwind the entanglement of the lines, as the examples 
in Figure \ref{fig:unwind}.  

\begin{figure}
\centering
\includegraphics[width=5cm]{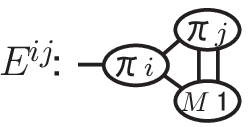}
\caption{The graphical representation of $E^{ij}\ (i,j=1,2)$.} 
\label{fig:eij}
\end{figure}

\begin{figure}
\centering
\includegraphics[width=15cm]{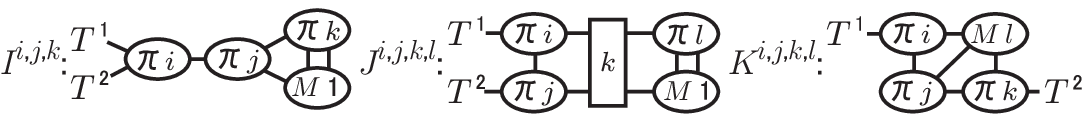}
\caption{The graphical representation of $I,J,K$.} 
\label{fig:ijk}
\end{figure}

By considering all the possible ways of the index contractions of the term represented by $G^4$,
one can find that four of them are independent and 
can be represented by a graph with two signatures, 
$E^{ij}\ (i,j=\pm 1)$, defined in Figure \ref{fig:eij}.   
The Hamiltonian constraint, which was ambiguously represented by $G^4$ in the previous sections, 
can generally have a form,
\be
{\cal H}_a = c_{i,j}E^{ij}_a,
\label{eq:haorder}
\ee
where $c_{i,j}\ (i,j=\pm 1)$ are real numerical coefficients. The computations of the Poisson
brackets of $E^{ij}_a$ can be carried out in the same manner as in the previous sections,
but this time the connections of the lines are precisely taken into account. Then one obtains
\be
\{ T^1_a E_a^{ij},T^2_b E^{kl}_b\}=I^{i,-jk,l}+J^{-i,-j,k,l}
+J^{-i,j,-k,l}
+J^{i,k,1,-jl}+K^{-i,-j,k,l}+K^{-i,j,k,-l}-(T^1ij\leftrightarrow T^2kl),
\label{eq:EE}
\ee
where $I,J,K$ are defined in Figure \ref{fig:ijk}.
It is easy to show that these $I,J,K$ are transformed by 
the interchange $T^1\leftrightarrow T^2$ to 
\beal
I^{i,j,k}&\rightarrow I^{-i,j,k}, \\
J^{i,j,k,l}&\rightarrow J^{-j,-i,-k,l}, \\
K^{i,j,k,l}&\rightarrow K^{-k,-j,-i,-l}.
\label{eq:ijktrans}
\eeal
Then, from \eq{eq:haorder}, \eq{eq:EE} and \eq{eq:ijktrans}, one obtains
\beal
\{& H(T^1),H(T^2)\}=
(c_{i,l}c_{-lj,k}-c_{-i,l}c_{-lj,k})I^{i,j,k}
+
(c_{-i,-j}c_{k,l}+c_{-i,j}c_{-k,l}-c_{j,i}c_{-k,l}-c_{j,-i}c_{k,l})J^{i,j,k,l}\\
&
+c_{i,l}c_{j,-kl} J^{i,j,1,k}-c_{-j,l}c_{-i,-kl}J^{i,j,-1,k}
+
(c_{-i,-j}c_{k,l}+c_{-i,j}c_{k,-l}-c_{k,j}c_{-i,-l}-c_{k,-j}c_{-i,l})K^{i,j,k,l}.
\label{eq:hhc}
\eeal

By comparing Figures \ref{fig:eij} and \ref{fig:ijk}, one finds that $c_{j,k}I^{i,j,k}$ 
contains the Hamiltonian constraint \eq{eq:haorder} as its part. 
And, from the expression \eq{eq:expd}, one sees also that $I^{i,1,-1}-I^{i,-1,-1}$ and 
$J^{i,j,1,-1}-J^{i,j,-1,-1}$ contain ${\cal D}_{ab}$. Therefore, from 
\eq{eq:hhc},
the condition for the algebraic closure of the constraint algebra is obtained as
\beal
&c_{i,l}c_{-lj,k}-c_{-i,l}c_{-lj,k}=\lambda_i c_{j,k} +\tilde \lambda_i (\delta_{j,1}\delta_{k,-1} 
 -\delta_{j,-1}\delta_{k,-1}),\\
&c_{-i,-j}c_{k,l}+c_{-i,j}c_{-k,l}-c_{j,i}c_{-k,l}-c_{j,-i}c_{k,l}
+c_{i,m}c_{j,-lm}\delta_{k,1} - c_{-j,m}c_{-i,-lm}\delta_{k,-1}\\
&\ \ \ \ \ \ \ \ \ \ \ \ \ \ \ \ \ \ \ \ \ 
\ \ \ \ \ \ \ \ \ \ \ \ \ \ \ \ \ \ \ \ \ 
\ \ \ \ \ \ \ \ \ \ \ \ \ \ \ \ \ \ \ \ \ =\lambda_{ij} \delta_{k,1}\delta_{l,-1}
-\lambda_{ij}\delta_{k,-1}\delta_{l,-1}, \\
&c_{-i,-j}c_{k,l}+c_{-i,j}c_{k,-l}-c_{k,j}c_{-i,-l}-c_{k,-j}c_{-i,l}=0,
\label{eq:ordercond}
\eeal
where $\lambda_i$, $\tilde \lambda_i$, and $\lambda_{ij}$ are the real numbers which express
the proportionality to the Hamiltonian constraint or ${\cal D}_{ab}$.

Because of the antisymmetry of the 
left-hand side of the first equation of \eq{eq:ordercond} under $i\rightarrow -i$, 
\be
(\lambda_i+\lambda_{-i}) c_{j,k} +(\tilde \lambda_i+\tilde \lambda_{-i}) (\delta_{j,1}\delta_{k,-1} 
 -\delta_{j,-1}\delta_{k,-1})=0
\label{eq:antilamdareq}
\ee
must be satisfied. Since a Hamiltonian constraint and ${\cal D}_{ab}$ should be independent
as a physical requirement, the former and the latter terms in \eq{eq:antilamdareq}
must vanish independently. Therefore
\beal
\lambda_{-i}&=-\lambda_i,\\
\tilde \lambda_{-i}&=-\tilde \lambda_i,
\label{eq:antilambda}
\eeal
can be assumed.

From the algebraic closure of ${\cal D}_{ab}$ expressed as \eq{eq:poissondd} and \eq{eq:poissondh},
it is obvious that one can freely add $\pi_{abc} {\cal D}_{bc}$, which can be expressed in
the form of the right-hand side of \eq{eq:haorder},
to a Hamiltonian constraint without violating the closure of a constraint algebra.
Therefore one can reduce the number of free parameters of the solutions to \eq{eq:ordercond} 
by fixing this free addition.
From \eq{eq:expd}, \eq{eq:haorder} and Figure \ref{fig:eij},
this addition corresponds to a shift of $c_{i,j}$ by
\be
\delta c_{i,j}=\epsilon(\delta_{i,1}\delta_{j,-1}-\delta_{i,-1}\delta_{j,-1}),
\label{eq:shiftc}
\ee
where $\epsilon$ is an infinitesimal parameter.
By substituting \eq{eq:shiftc} into the first equation of \eq{eq:ordercond}, one finds that
\eq{eq:shiftc} is equivalent to\footnote{$\tilde \lambda_i$ is also shifted.}
\be
\delta \lambda_i=2\epsilon( \delta_{i,1}- \delta_{i,-1}).
\label{eq:shiftlambda}
\ee
Therefore, from \eq{eq:antilambda} and \eq{eq:shiftlambda}, one can safely assume 
\be
\lambda_i=0
\ee
without loss of generality, with the allowance that one may freely add \eq{eq:shiftc} 
to the solutions.

Then the first equation of \eq{eq:ordercond} gives
\beal
&[(c_{1,1}-c_{-1,1})+(c_{1,-1}-c_{-1,-1})](c_{1,k}+c_{-1,k})=0,\\
&(c_{1,1}-c_{-1,1})c_{-j,1}+(c_{1,-1}-c_{-1,-1})c_{j,1}=0.
\label{eq:first2}
\eeal
Here the first equation of \eq{eq:first2} has been derived by summing over $j=\pm 1$
to delete $\tilde \lambda_i$, and
the second of \eq{eq:first2} by substituting $k=1$.  
The second equation of \eq{eq:first2} can be used to delete $c_{1,-1}-c_{-1,-1}$ from the
first equation of \eq{eq:first2} to obtain
\be
(c_{1,1}-c_{-1,1})^2(c_{1,k}+c_{-1,k})=0.
\ee
This shows that $c_{1,1}=c_{-1,1}$ or $c_{1,k}=-c_{-1,k}\ (k=\pm1)$. Substituting the first case
into \eq{eq:first2} leads to $c_{1,-1}=c_{-1,-1}$ or $c_{i1}=0$ and 
$(c_{1,-1}-c_{-1,-1})(c_{1,-1}+c_{-1,-1})=0$. Thus the solutions to the first equation
of \eq{eq:ordercond} can be classified into the following two cases,
\beal
&\hbox{(i)}\ c_{1,k}=c_{-1,k}\ (k=\pm1),\\
&\hbox{(ii)}\ c_{1,k}=-c_{-1,k}\ (k=\pm1).
\label{eq:iandii}
\eeal

By substituting the case (i) of \eq{eq:iandii} into the second equation of \eq{eq:ordercond}, one 
obtains  
\be
c_{1,1} c_{1,-1}=0.
\ee
Therefore there are two non-vanishing solutions, 
\beal
&\hbox{(i-1) }c_{1,1}=c_{-1,1}\neq 0,\ c_{1,-1}=c_{-1,-1}=0, \\
&\hbox{(i-2) }c_{1,1}=c_{-1,1}=0,\ c_{1,-1}=c_{-1,-1}\neq 0.
\label{eq:finalsolution}
\eeal
On the other hand, 
by substituting the case (ii) of \eq{eq:iandii} into the second equation of \eq{eq:ordercond}, one 
obtains
\be
(c_{1,1})^2=0.
\ee
The solution, $c_{1,1}=c_{-1,1}=0,c_{1,-1}=-c_{-1,-1}$, just represents $\pi_{abc}{\cal D}_{bc}$, 
and is physically rejected as a Hamiltonian constraint.
The third equation of \eq{eq:ordercond} is always satisfied. 
It can be checked that the two solutions in \eq{eq:finalsolution} actually satisfy all the equations
in \eq{eq:ordercond}.

Thus the conclusion is that, from \eq{eq:finalsolution}, 
the forms of a Hamiltonian constraint which satisfy the algebraic consistency 
are exhausted by
\beal
{\cal H}_a&=\pi_{a(bc)}\pi_{bde}M_{cde},\\
{\cal H}_a&=\pi_{a(bc)}\pi_{bde}M_{ced},
\label{eq:finalham}
\eeal
where $(\ )$ denotes symmetrization of the indices, $\pi_{a(bc)}\equiv (\pi_{abc}+\pi_{acb})/2$.

In the both cases of \eq{eq:finalham}, one obtains,
\be
\{H(T^1),H(T^2)\}=D([\tilde T^1,\tilde T^2]),
\label{eq:finalhh}
\ee
where $\tilde T^i$ are the symmetric matrices defined by
\be
\tilde T^i{}_{bc}=T^i_a \pi_{a(bc)}.
\label{eq:defttilde}
\ee 
As was discussed in the previous paper \cite{Sasakura:2011sq}, the 
constraint algebra \eq{eq:poissondd}, 
\eq{eq:poissondh} and \eq{eq:finalhh} would reproduce the first class 
constraint algebra of the canonical
formalism of the general relativity for the Minkowski signature in the pointwise 
limit\footnote{It seems impossible
to reproduce the first class constraint algebra of general relativity in the Euclidean signature, 
though there are no proofs for no-go.}. 
In the derivation, the dependence of $\tilde T^i$ on the canonical variables
as in \eq{eq:defttilde} played the essential roles.

Let me next discuss the possibility of adding a term represented by $G^1$ to ${\cal H}_a$.
The Poisson bracket of Hamiltonian constraints will be changed by
\be
\{H(T^1),T^2_a M_{abb}\}+\{T^1_a M_{abb},H(T^2)\}=-2 (T^1_a T^2_b-T^1_bT^2_a)
\pi_{a(cd)}M_{b(cd)}.
\label{eq:costerm}
\ee
The right-hand side looks very similar to but is not the same as ${\cal D}_{ab}$ 
because of the symmetrization of the contracted indices.
Therefore the algebra will not close, and one cannot add the term represented by $G^1$ to ${\cal H}_a$.

In the previous paper, another form of a Hamiltonian constraint has been presented. 
Indeed one can easily check 
that the previous form, except for the linear term,
can be obtained by applying a canonical transformation 
$M\rightarrow (M\pm \pi)\sqrt{2}$, $\pi \rightarrow (\pi\mp M)/\sqrt{2}$ to the 
first one in \eq{eq:finalham}. The absence of the linear term is 
because it corresponds to the term $\pi_{abb}$ in the parameterization of this paper, and 
has been rejected due to its violation of the time reversal symmetry. 

\section{Summary and future prospects}
\label{sec:summary}
In the previous paper \cite{Sasakura:2011sq}, 
Hamiltonian formalism of the rank-three tensor model has been proposed,
in which ``local" time is consistently incorporated by a first class constraint algebra.
A consistent Hamiltonian constraint was presented in the paper, but it was not clear whether there 
were other possibilities or not. Moreover, it behaved oddly under the time reversal transformation.

To solve these issues, 
this paper has given the thorough discussions on 
the allowed forms of a Hamiltonian constraint, assuming the 
physically reasonable conditions (i)-(vi) listed in Section \ref{sec:introduction},
among which (iv) imposes the time reversal symmetry. 
The closure condition of a constraint algebra strongly restricts the allowed forms, and 
it has been shown that there exist only two consistent forms, which have a slight difference
in index contractions. 
The Hamiltonian constraint obtained in the previous paper 
can indeed be transformed to one of them by a canonical change of variables. 

One must be cautious about how far the two-fold uniqueness has been shown in this paper. 
In fact, in passing from the unordered computations to the ordered ones, 
the possibilities for certain linear combinations of terms to exist in a Hamiltonian constraint have been ignored. 
To explain more concretely, let me consider the ordered graphs $\tilde G^{\alpha\, i}$,
where $\alpha$ is a new label,
corresponding to the unordered graphs $G^i$ in Figure \ref{fig:possible}.
Then, since the unordered computations do not recognize the difference between $\tilde G^{\alpha\, i}$ and 
$\tilde G^{\beta\, i}$, the unordered computations of Poisson brackets remain the same even if a linear combination 
$\sum_{\alpha,i} \tilde c_{\alpha\,i} \tilde G^{\alpha\, i}$ 
with $\sum_\alpha \tilde c_{\alpha\, i}=0$ is added to a Hamiltonian
constraint.
This means that it is not right in Section \ref{sec:order} to assume that there only exist the ordered graphs corresponding to 
$G^4$ in a Hamiltonian constraint, based on the unordered computations. 
In this sense, a complete proof of the two-fold uniqueness has not been given in this paper.

On the other hand, from the prospective of gravitational physics, 
the uniqueness has been proven.
As discussed in \cite{Sasakura:2006pq}-\cite{Sasakura:2010rb}, the gravitational degrees of freedom 
are described by the commutative part of fuzzy spaces. This part corresponds to the totally symmetric part of the tensors.
Therefore, for the purpose of quantum gravity, one could formulate the rank-three tensor models 
with totally symmetric real tensors instead of Hermitian complex-valued tensors of this paper.  
Then the unordered computations are enough and a Hamiltonian constraint can only have $G^4$ and $G^1$.
For totally symmetric tensors, the two final forms in \eq{eq:finalham} are identical, and the uniqueness is literally true. 
Another new thing is that the right-hand side of \eq{eq:costerm} is proportional to ${\cal D}_{ab}$, and therefore
$G^1$ is also allowed to exist in a Hamiltonian constraint. 
 
Once a dynamical system is defined by a first class constraint algebra, the next future questions would be 
how to quantize it and what is its dynamics. In the present motivation for 
the rank-three tensor model, 
highly interesting would be the question of whether there exist the classical regimes in 
which the system behaves as if there is a spacetime respecting locality - {\it emergent spacetime}
\cite{Sindoni:2011ej}.
The two-fold uniqueness of the Hamiltonian constraint shown in this paper will 
simplify the future study of this question in the rank-three tensor model.

%\section*{Acknowledgement}

%\bibliography{unique}

\begin{thebibliography}{99}

%\cite{Garay:1994en}
\bibitem{Garay:1994en}
  L.~J.~Garay,
  ``Quantum gravity and minimum length,''
  Int.\ J.\ Mod.\ Phys.\  A {\bf 10}, 145 (1995)
  [arXiv:gr-qc/9403008].
  %%CITATION = IMPAE,A10,145;%%
  

%\cite{Sindoni:2011ej}
\bibitem{Sindoni:2011ej} 
  L.~Sindoni,
  ``Emergent models for gravity: An Overview,''  arXiv:1110.0686 [gr-qc].  %%CITATION = ARXIV:1110.0686;%%


%\cite{Connes:1994yd}
\bibitem{Connes:1994yd} 
  A.~Connes,
  ``Noncommutative geometry,''  %%CITATION = INSPIRE-391003;%%

%\cite{Madore:1991bw}
\bibitem{Madore:1991bw}
  J.~Madore,
  ``The Fuzzy sphere,''
  Class.\ Quant.\ Grav.\  {\bf 9}, 69-88 (1992).


 %\cite{deMedeiros:2004wb}
\bibitem{deMedeiros:2004wb}
  P.~de Medeiros and S.~Ramgoolam,
  ``Non-associative gauge theory and higher spin interactions,''
  JHEP {\bf 0503}, 072 (2005)
  [arXiv:hep-th/0412027].
  %%CITATION = JHEPA,0503,072;%%

%\cite{Ramgoolam:2003cs}
\bibitem{Ramgoolam:2003cs}
  S.~Ramgoolam,
  ``Towards gauge theory for a class of commutative and nonassociative fuzzy
  spaces,''
  JHEP {\bf 0403}, 034 (2004)
  [arXiv:hep-th/0310153].
  %%CITATION = JHEPA,0403,034;%%

%\cite{Ramgoolam:2001zx}
\bibitem{Ramgoolam:2001zx}
  S.~Ramgoolam,
  ``On spherical harmonics for fuzzy spheres in diverse dimensions,''
  Nucl.\ Phys.\  B {\bf 610}, 461 (2001)
  [arXiv:hep-th/0105006].
  %%CITATION = NUPHA,B610,461;%%
  
%\cite{Sasai:2006ua}
\bibitem{Sasai:2006ua}
  Y.~Sasai, N.~Sasakura,
  ``One-loop unitarity of scalar field theories on Poincare invariant commutative nonassociative spacetimes,''
  JHEP {\bf 0609}, 046 (2006).
  [hep-th/0604194].

%\cite{Sasakura:2005js}
\bibitem{Sasakura:2005js} 
  N.~Sasakura,
  ``An Invariant approach to dynamical fuzzy spaces with a three-index variable,''  Mod.\ Phys.\ Lett.\ A {\bf 21}, 1017 (2006)  [hep-th/0506192].  %%CITATION = HEP-TH/0506192;%%

%\cite{Sasakura:2011ma}
\bibitem{Sasakura:2011ma} 
  N.~Sasakura,
  ``Tensor models and 3-ary algebras,'' J.\ Math.\ Phys. {\bf 52}, 103510 (2011) [arXiv:1104.1463 [hep-th]].  %%CITATION = ARXIV:1104.1463;%%

%\cite{Sasakura:2011nj}
\bibitem{Sasakura:2011nj} 
  N.~Sasakura,
  ``Tensor models and hierarchy of n-ary algebras,''  Int.\ J.\ Mod.\ Phys.\ A {\bf 26}, 3249 (2011)  [arXiv:1104.5312 [hep-th]].  %%CITATION = ARXIV:1104.5312;%%

%\cite{Sasakura:2011qg}
\bibitem{Sasakura:2011qg} 
  N.~Sasakura,
  ``Super tensor models, super fuzzy spaces and super n-ary transformations,''  Int.\ J.\ Mod.\ Phys.\ A {\bf 26}, 4203 (2011)  [arXiv:1106.0379 [hep-th]].  %%CITATION = ARXIV:1106.0379;%%

%\cite{Ambjorn:1990ge}
\bibitem{Ambjorn:1990ge}
  J.~Ambjorn, B.~Durhuus and T.~Jonsson,
  ``Three-Dimensional Simplicial Quantum Gravity And Generalized Matrix
  Models,''
  Mod.\ Phys.\ Lett.\ A {\bf 6}, 1133 (1991).
  %%CITATION = MPLAE,A6,1133;%%

%\cite{Sasakura:1990fs}
\bibitem{Sasakura:1990fs}
  N.~Sasakura,
  ``Tensor Model For Gravity And Orientability Of Manifold,''
  Mod.\ Phys.\ Lett.\ A {\bf 6}, 2613 (1991).
  %%CITATION = MPLAE,A6,2613;%%

%\cite{Godfrey:1990dt}
\bibitem{Godfrey:1990dt}
  N.~Godfrey and M.~Gross,
  ``Simplicial Quantum Gravity In More Than Two-Dimensions,''
  Phys.\ Rev.\ D {\bf 43}, 1749 (1991).
  %%CITATION = PHRVA,D43,1749;%%
  
%\cite{Boulatov:1992vp}
\bibitem{Boulatov:1992vp}
  D.~V.~Boulatov,
  ``A Model of three-dimensional lattice gravity,''
  Mod.\ Phys.\ Lett.\ A {\bf 7}, 1629 (1992)
  [arXiv:hep-th/9202074].
  %%CITATION = HEP-TH 9202074;%%

%\cite{Ooguri:1992eb}
\bibitem{Ooguri:1992eb}
  H.~Ooguri,
  ``Topological lattice models in four-dimensions,''
  Mod.\ Phys.\ Lett.\ A {\bf 7}, 2799 (1992)
  [arXiv:hep-th/9205090].
  %%CITATION = HEP-TH 9205090;%%

%\cite{DePietri:1999bx}
\bibitem{DePietri:1999bx}
  R.~De Pietri, L.~Freidel, K.~Krasnov and C.~Rovelli,
  ``Barrett-Crane model from a Boulatov-Ooguri field theory over a  homogeneous
  space,''
  Nucl.\ Phys.\ B {\bf 574}, 785 (2000)
  [arXiv:hep-th/9907154].
  %%CITATION = HEP-TH 9907154;%%   
  
%\cite{Freidel:2005qe}
\bibitem{Freidel:2005qe} 
  L.~Freidel,
  ``Group field theory: An Overview,''  Int.\ J.\ Theor.\ Phys.\  {\bf 44}, 1769 (2005)  [hep-th/0505016].  %%CITATION = HEP-TH/0505016;%%

%\cite{Oriti:2011jm}
\bibitem{Oriti:2011jm} 
  D.~Oriti,
  ``The microscopic dynamics of quantum space as a group field theory,''  arXiv:1110.5606 [hep-th].  %%CITATION = ARXIV:1110.5606;%%

%\cite{Gurau:2009tw}
\bibitem{Gurau:2009tw} 
  R.~Gurau,
  ``Colored Group Field Theory,''  Commun.\ Math.\ Phys.\  {\bf 304}, 69 (2011)  [arXiv:0907.2582 [hep-th]].  %%CITATION = ARXIV:0907.2582;%%

%\cite{Bonzom:2012hw}
\bibitem{Bonzom:2012hw} 
  V.~Bonzom, R.~Gurau and V.~Rivasseau,
  ``Random tensor models in the large N limit: Uncoloring the colored tensor models,''  arXiv:1202.3637 [hep-th].  %%CITATION = ARXIV:1202.3637;%%

%\cite{Bonzom:2012sz}
\bibitem{Bonzom:2012sz} 
  V.~Bonzom,
  ``Multicritical tensor models and hard dimers on spherical random lattices,''  arXiv:1201.1931 [hep-th].  %%CITATION = ARXIV:1201.1931;%%

 %\cite{BenGeloun:2012pu}
\bibitem{BenGeloun:2012pu} 
  J.~Ben Geloun and D.~O.~Samary,
  ``3D Tensor Field Theory: Renormalization and One-loop $\beta$-functions,''  arXiv:1201.0176 [hep-th].  %%CITATION = ARXIV:1201.0176;%%

 %\cite{Rivasseau:2011hm}
\bibitem{Rivasseau:2011hm} 
  V.~Rivasseau,
  ``Quantum Gravity and Renormalization: The Tensor Track,''  arXiv:1112.5104 [hep-th].  %%CITATION = ARXIV:1112.5104;%%

 %\cite{Baratin:2011aa}
\bibitem{Baratin:2011aa} 
  A.~Baratin and D.~Oriti,
  ``Ten questions on Group Field Theory (and their tentative answers),''  arXiv:1112.3270 [gr-qc].  %%CITATION = ARXIV:1112.3270;%%

 %\cite{BenGeloun:2011rc}
\bibitem{BenGeloun:2011rc} 
  J.~Ben Geloun and V.~Rivasseau,
  ``A Renormalizable 4-Dimensional Tensor Field Theory,''  arXiv:1111.4997 [hep-th].  %%CITATION = ARXIV:1111.4997;%%

 %\cite{Gurau:2011sk}
\bibitem{Gurau:2011sk} 
  R.~Gurau,
  ``The Double Scaling Limit in Arbitrary Dimensions: A Toy Model,''  Phys.\ Rev.\ D {\bf 84}, 124051 (2011)  [arXiv:1110.2460 [hep-th]].  %%CITATION = ARXIV:1110.2460;%%

 %\cite{Bonzom:2011jv}
\bibitem{Bonzom:2011jv} 
  V.~Bonzom and A.~Laddha,
  ``Lessons from toy-models for the dynamics of loop quantum gravity,''  arXiv:1110.2157 [gr-qc].  %%CITATION = ARXIV:1110.2157;%%

 %\cite{Gurau:2011xp}
\bibitem{Gurau:2011xp} 
  R.~Gurau and J.~P.~Ryan,
  ``Colored Tensor Models - a review,''  arXiv:1109.4812 [hep-th].  %%CITATION = ARXIV:1109.4812;%%

 %\cite{Bonzom:2011ev}
\bibitem{Bonzom:2011ev} 
  V.~Bonzom, R.~Gurau and V.~Rivasseau,
  ``The Ising Model on Random Lattices in Arbitrary Dimensions,''  arXiv:1108.6269 [hep-th].  %%CITATION = ARXIV:1108.6269;%%

 %\cite{Benedetti:2011nn}
\bibitem{Benedetti:2011nn} 
  D.~Benedetti and R.~Gurau,
  ``Phase Transition in Dually Weighted Colored Tensor Models,''  Nucl.\ Phys.\ B {\bf 855}, 420 (2012)  [arXiv:1108.5389 [hep-th]].  %%CITATION = ARXIV:1108.5389;%%

 %\cite{Baratin:2011tx}
\bibitem{Baratin:2011tx} 
  A.~Baratin and D.~Oriti,
  ``Quantum simplicial geometry in the group field theory formalism: reconsidering the Barrett-Crane model,''  New J.\ Phys.\  {\bf 13}, 125011 (2011)  [arXiv:1108.1178 [gr-qc]].  %%CITATION = ARXIV:1108.1178;%%

 %\cite{Gurau:2011tj}
\bibitem{Gurau:2011tj} 
  R.~Gurau,
  ``A generalization of the Virasoro algebra to arbitrary dimensions,''  Nucl.\ Phys.\ B {\bf 852}, 592 (2011)  [arXiv:1105.6072 [hep-th]].  %%CITATION = ARXIV:1105.6072;%%

 %\cite{Bonzom:2011zz}
\bibitem{Bonzom:2011zz} 
  V.~Bonzom, R.~Gurau, A.~Riello and V.~Rivasseau,
  ``Critical behavior of colored tensor models in the large N limit,''  Nucl.\ Phys.\ B {\bf 853}, 174 (2011)  [arXiv:1105.3122 [hep-th]].  %%CITATION = ARXIV:1105.3122;%%

 %\cite{Livine:2011yb}
\bibitem{Livine:2011yb} 
  E.~R.~Livine, D.~Oriti and J.~P.~Ryan,
  ``Effective Hamiltonian Constraint from Group Field Theory,''  Class.\ Quant.\ Grav.\  {\bf 28}, 245010 (2011)  [arXiv:1104.5509 [gr-qc]].  %%CITATION = ARXIV:1104.5509;%%

%\cite{Carrozza:2011jn}
\bibitem{Carrozza:2011jn} 
  S.~Carrozza and D.~Oriti,
  ``Bounding bubbles: the vertex representation of 3d Group Field Theory and the suppression of pseudo-manifolds,''  Phys.\ Rev.\ D {\bf 85}, 044004 (2012)  [arXiv:1104.5158 [hep-th]].  %%CITATION = ARXIV:1104.5158;%%

%\cite{Gurau:2011xq}
\bibitem{Gurau:2011xq} 
  R.~Gurau,
  ``The complete 1/N expansion of colored tensor models in arbitrary dimension,''  arXiv:1102.5759 [gr-qc].  %%CITATION = ARXIV:1102.5759;%%

%\cite{Gurau:2010ba}
\bibitem{Gurau:2010ba} 
  R.~Gurau,
  ``The 1/N expansion of colored tensor models,''  Annales Henri Poincare {\bf 12}, 829 (2011)  [arXiv:1011.2726 [gr-qc]].  %%CITATION = ARXIV:1011.2726;%%

 %\cite{Geloun:2010vj}
\bibitem{Geloun:2010vj} 
  J.~Ben Geloun, R.~Gurau and V.~Rivasseau,
  ``EPRL/FK Group Field Theory,''  Europhys.\ Lett.\  {\bf 92}, 60008 (2010)  [arXiv:1008.0354 [hep-th]].  %%CITATION = ARXIV:1008.0354;%%

 %\cite{Gurau:2010nd}
\bibitem{Gurau:2010nd} 
  R.~Gurau,
  ``Lost in Translation: Topological Singularities in Group Field Theory,''  Class.\ Quant.\ Grav.\  {\bf 27}, 235023 (2010)  [arXiv:1006.0714 [hep-th]].  %%CITATION = ARXIV:1006.0714;%%

%\cite{Gurau:2009tz}
\bibitem{Gurau:2009tz} 
  R.~Gurau,
  ``Topological Graph Polynomials in Colored Group Field Theory,''  Annales Henri Poincare {\bf 11}, 565 (2010)  [arXiv:0911.1945 [hep-th]].  %%CITATION = ARXIV:0911.1945;%%

%\cite{Sasakura:2006pq}
\bibitem{Sasakura:2006pq} 
  N.~Sasakura,
  ``Tensor model and dynamical generation of commutative nonassociative fuzzy spaces,''  Class.\ Quant.\ Grav.\  {\bf 23}, 5397 (2006)  [hep-th/0606066].  %%CITATION = HEP-TH/0606066;%%

%\cite{Sasakura:2007sv}
\bibitem{Sasakura:2007sv} 
  N.~Sasakura,
  ``The Fluctuation spectra around a Gaussian classical solution of a tensor model and the general relativity,''  Int.\ J.\ Mod.\ Phys.\ A {\bf 23}, 693 (2008)  [arXiv:0706.1618 [hep-th]].  %%CITATION = ARXIV:0706.1618;%%

%\cite{Sasakura:2007ud}
\bibitem{Sasakura:2007ud} 
  N.~Sasakura,
  ``The Lowest modes around Gaussian solutions of tensor models and the general relativity,''  Int.\ J.\ Mod.\ Phys.\ A {\bf 23}, 3863 (2008)  [arXiv:0710.0696 [hep-th]].  %%CITATION = ARXIV:0710.0696;%%

%\cite{Sasakura:2008pe}
\bibitem{Sasakura:2008pe} 
  N.~Sasakura,
  ``Emergent general relativity on fuzzy spaces from tensor models,''  Prog.\ Theor.\ Phys.\  {\bf 119}, 1029 (2008)  [arXiv:0803.1717 [gr-qc]].  %%CITATION = ARXIV:0803.1717;%%

%\cite{Sasakura:2009hs}
\bibitem{Sasakura:2009hs} 
  N.~Sasakura,
  ``Gauge fixing in the tensor model and emergence of local gauge symmetries,''  Prog.\ Theor.\ Phys.\  {\bf 122}, 309 (2009)  [arXiv:0904.0046 [hep-th]].  %%CITATION = ARXIV:0904.0046;%%

%\cite{Sasakura:2010rb}
\bibitem{Sasakura:2010rb} 
  N.~Sasakura,
  ``A Renormalization procedure for tensor models and scalar-tensor theories of gravity,''  Int.\ J.\ Mod.\ Phys.\ A {\bf 25}, 4475 (2010)  [arXiv:1005.3088 [hep-th]].  %%CITATION = ARXIV:1005.3088;%%

%\cite{Ambjorn:2010ce}
\bibitem{Ambjorn:2010ce} 
  J.~Ambjorn, A.~Gorlich, J.~Jurkiewicz and R.~Loll,
  ``CDT - an Entropic Theory of Quantum Gravity,''  arXiv:1007.2560 [hep-th].  %%CITATION = ARXIV:1007.2560;%%

%\cite{Geloun:2011cy}
\bibitem{Geloun:2011cy} 
  J.~Ben Geloun and V.~Bonzom,
  ``Radiative corrections in the Boulatov-Ooguri tensor model: The 2-point function,''  Int.\ J.\ Theor.\ Phys.\  {\bf 50}, 2819 (2011)  [arXiv:1101.4294 [hep-th]].  %%CITATION = ARXIV:1101.4294;%%

%\cite{Sasakura:2011sq}
\bibitem{Sasakura:2011sq} 
  N.~Sasakura,
  ``Canonical tensor models with local time,'' Int.\ J.\ Mod.\ Phys.\ A {\bf 27}, 1250020 (2012) 
  [arXiv:1111.2790 [hep-th]].  %%CITATION = ARXIV:1111.2790;%%

%\cite{Arnowitt:1962hi}
\bibitem{Arnowitt:1962hi} 
  R.~L.~Arnowitt, S.~Deser and C.~W.~Misner,
  ``The Dynamics of general relativity,''  gr-qc/0405109.  %%CITATION = GR-QC/0405109;%%

%\cite{DeWitt:1967yk}
\bibitem{DeWitt:1967yk} 
  B.~S.~DeWitt,
  ``Quantum Theory of Gravity. 1. The Canonical Theory,''  Phys.\ Rev.\  {\bf 160}, 1113 (1967).  %%CITATION = PHRVA,160,1113;%%

%\cite{Hojman:1976vp}
\bibitem{Hojman:1976vp} 
  S.~A.~Hojman, K.~Kuchar and C.~Teitelboim,
  ``Geometrodynamics Regained,''  Annals Phys.\  {\bf 96}, 88 (1976).  %%CITATION = APNYA,96,88;%%

\end{thebibliography}

\end{document}